\documentclass[preprint2]{aastex}
\usepackage{amsmath}
\usepackage{natbib}

\begin{document}
\title
{
  A new luminosity relation for gamma-ray bursts and its implication
}
\author
{
  Shi Qi$^{1}$
  and
  Tan Lu$^{2}$
}
\affil
{
  Purple Mountain Observatory, Chinese Academy of Sciences, Nanjing
  210008, China
}
\affil
{
  Joint Center for Particle, Nuclear Physics and Cosmology, Nanjing
  University --- Purple Mountain Observatory, Nanjing  210093, China
}
\email{$^{1}$qishi11@gmail.com}
\email{$^{2}$t.lu@pmo.ac.cn}

\begin{abstract}
  Gamma-ray bursts (GRBs) are the most luminous astrophysical events
  observed so far.
  They are conventionally classified into long and short ones
  depending on their time duration, $T_{90}$.
  Because of the advantage their high redshifts offer, many efforts
  have been made to apply GRBs to cosmology.
  The key to this is to find correlations between some measurable
  properties of GRBs and the energy or the luminosity of GRBs.
  These correlations are usually referred to as luminosity relations
  and are helpful in understanding the GRBs themselves.
  In this paper, we explored such correlations in the X-ray emission
  of GRBs.
  The X-ray emission of GRBs observed by \emph{Swift} has the
  exponential functional form in the prompt phase and relaxes to a
  power-law decay at time $T_p$.
  We have assumed a linear relation between $\log L_{X, p}$ (with
  $L_{X, p}$ being the X-ray luminosity at $T_p$)
  and $\log [T_p/(1+z)]$,
  but there is some evidence for curvature in the data and the true
  relationship between $L_{X, p}$ and $T_p/(1+z)$ may be a broken
  power law.
  The limited GRB sample used in our analysis is still not sufficient
  for us to conclude whether the break is real or just an illusion
  caused by outliers.
  We considered both cases in our analysis and discussed the
  implications of the luminosity relation, especially on the time
  duration of GRBs and their classification.
\end{abstract}


\section{Introduction}

Gamma-ray bursts (GRBs), which can last from milliseconds to nearly an
hour, are the most luminous astrophysical events observed so far.
The parameter of $T_{90}$, which is defined as the time interval
during which the background-subtracted cumulative counts increase from
$5\%$ to $95\%$, is usually used to denote the time duration of GRBs.
Those with $T_{90} > 2 \, \mathrm{s}$ are conventionally described as
long/soft GRBs and those with $T_{90} < 2 \, \mathrm{s}$ as short/hard
GRBs~\citep{Kouveliotou:1993yx}.

In~\citet{Willingale:2006zh}, it was demonstrated that the X-ray decay
curves of GRBs measured by \emph{Swift} can be
fitted using one or two components---the prompt component and the
optional afterglow component---both of which have exactly the same
functional form comprised of an early falling exponential phase
and a following power-law decay. The prompt component contains the
prompt gamma-ray emission and the initial X-ray decay. The transition
time $T_p$ from the exponential phase to the power-law decay in the
prompt component defines an alternative estimate of the GRB duration,
which is comparable with $T_{90}$ for most
GRBs~\citep{O'Brien:2006sa, O'Brien:2006jz}.
\citet{O'Brien:2007apss} proposed the classification of GRBs into
long and short ones at $T_p = 10 \, \mathrm{s}$ instead.

GRBs can be observed at very high redshifts due to their high
luminosities. For example, the recently observed GRB 090423 has a
redshift of
$z \approx 8.2$~\citep{Tanvir:2009ex, Salvaterra:2009ey}.
It may be possible to calibrate GRBs as
standard candles~\citep[see, for example,][etc.]{Dai:2004tq,
  Ghirlanda:2004fs, Firmani:2005gs, Lamb:2005cw, Liang:2005xb,
  Xu:2005uv, Wang:2005ic, Ghirlanda:2006ax, Schaefer:2006pa,
  Wang:2007rz, Li:2007re, Amati:2008hq, Basilakos:2008tp,
  Qi:2008zk, Qi:2008ag, Kodama:2008dq, Liang:2008kx, Wang:2008vja,
  Qi:2009yr}.
The key to the calibration of GRBs is to find correlations between
some measurable properties (the luminosity indicators) of GRBs and the
energy (the isotropic energy $E_{\gamma, \mathrm{iso}}$ or the
collimation-corrected energy $E_{\gamma}$) or the luminosity (e.g.,
the peak luminosity $L$) of GRBs.
These correlations are usually referred to as luminosity relations,
which are useful both in applying GRBs to cosmology and understanding
GRBs themselves.
The GRB luminosity relations used in cosmological studies
in the literature include the relations of
$\tau_{\mathrm{lag}}$ (the spectral lag, i.e., the time shift between
the hard and soft light curves)--$L$~\citep{Norris:1999ni},
$V$ (the variability, a quantitative measurement on the spikiness of
the light curve; there exist several definitions of $V$, mainly
depending on the smoothing time intervals the reference curve is built
upon, and on the normalization as well)--$L$~\citep{Fenimore:2000vs,
  Reichart:2000kq},
$E_{\mathrm{peak}}$ (peak energy
of the spectrum)--$E_{\gamma, \mathrm{iso}}$~\citep{Amati:2002ny},
$E_{\mathrm{peak}}$--$E_{\gamma}$~\citep{Ghirlanda:2004me},
$E_{\mathrm{peak}}$--$L$~\citep{Schaefer:2002tf},
and
$\tau_{\mathrm{RT}}$ (minimum rise time of
the light curve)--$L$~\citep{Schaefer:2006pa}.
For each of the above luminosity relations, there is only one
luminosity indicator involved.
More complicated luminosity relations, which include two luminosity
indicators are also discussed in the literature; see, for
example, \citet{Yu:2009xd} and references therein.

In this paper, we explore the correlation between $T_p$ and the
X-ray luminosity of GRBs at $T_p$ and discuss its implications,
especially on the time duration of GRBs and their classification.

\section{Methodology}

In~\citet{Willingale:2006zh}, the X-ray light curves of GRBs were
constructed from the combination of Burst Alert Telescope (BAT) and
X-ray telescope (XRT) data in the way
described by~\citet[the BAT data is extrapolated to the XRT
band]{O'Brien:2006jz}
and fitted using one or two
components---the prompt component and the optional afterglow
component---both of which have the same functional form:
\begin{equation}
  \label{eq:func_form}
  f_c(t)
  =
  \left\{
    \begin{array}{ll}
      F_c
      \exp \left( \alpha_c - \frac{t \alpha_c}{T_c} \right)
      \exp \left( \frac{-t_c}{t} \right)
      ,
      &
      t < T_c
      ,
      \\
      F_c
      \left( \frac{t}{T_c} \right)^{-\alpha_c}
      \exp \left( \frac{-t_c}{t} \right)
      ,
      &
      t \ge T_c
      .
    \end{array}
  \right.
\end{equation}
The transition from the exponential to the power law occurs at the
point $(T_c, \  F_c)$.
The subscript $c$ is replaced by $p$ for the prompt component and by
$a$ for the afterglow component, and the fitted X-ray flux is the sum
of the two components, i.e., $f(t) = f_p(t) + f_a(t)$.
In the derivation of the parameters, an initial fit was done to
find the peak position of the prompt emission; the peak time was then
used as time zero. A second fit was done with $t_p = 0$.
Thus, the derived parameters are all referenced with respect to the
estimated peak time rather than the somewhat arbitrary BAT trigger
time.
In addition, large flares have been masked out of the fitting
procedure~\citep[see][for a discussion on the average luminosity of
X-ray flares as a function of time]{Lazzati:2008da}.

We investigated the correlation between the transition time $T_p$ for
the prompt component and the X-ray luminosity of GRBs at
$T_p$~\citep[see also][for a discussion on the correlation between
$T_a$ and the luminosity at $T_a$]{Dainotti:2008vw}.
The fit of the GRB light curves directly gives the values of $T_p$.
The X-ray luminosity of GRBs at a given time $t$ is calculated by
\begin{equation}
  \label{eq:luminosity_X}
  L_X(t) = 4 \pi d_L^2(z) F_X(t)
  ,
\end{equation}
where $d_L(z)$ is the luminosity distance, for which we have used
the flat $\Lambda$CDM cosmological model with $\Omega_m=0.27$
and $H_0 = 71 \ \mathrm{km} \, \mathrm{s}^{-1} \, \mathrm{Mpc}^{-1}$,
and $F_X(t)$ is the $K$-corrected flux given by
\begin{align}
  \label{eq:flux_X}
  F_X(t)
  &=
  f(t)
  \times
  \frac
  {
    \int_{ E_{\mathrm{min}} / (1+z) }^{ E_{\mathrm{max}} / (1+z) }
    E^{-\beta}
    \mathrm{d} E
  }
  {
    \int_{ E_{\mathrm{min}} }^{ E_{\mathrm{max}} }
    E^{-\beta}
    \mathrm{d} E
  }
  \nonumber
  \\
  &=
  f(t)
  \times
  (1+z)^{\beta - 1}
  ,
\end{align}
where $\beta$ is the spectral index and, in general, it is time
dependent. \citet{Willingale:2006zh} presented in Table 3 of their
paper the spectral index
in the prompt phase ($\beta_p$; from the BAT data),
in the prompt decay ($\beta_{pd}$; from the XRT data),
on the plateau of the afterglow component ($\beta_a$; XRT data),
and in the final decay ($\beta_{ad}$; XRT data).
Since the BAT data have been extrapolated to the XRT band in the
combination of the BAT data and XRT data,
$(E_{\mathrm{min}}, \ E_{\mathrm{max}})
= (0.3, \ 10) \ \mathrm{keV}$
should be used here
and for this limited energy range, the simple power-law $E^{-\beta}$
is sufficient for the GRB spectrum.
So, the X-ray luminosity of GRBs at the transition time $T_p$ is
\begin{align}
  \label{eq:luminosity_Xp}
  L_{X, p} &= 4 \pi d_L^2(z) F_{X, p}
  \nonumber
  \\
  &=
  4 \pi d_L^2(z) F_p (1+z)^{\beta_p - 1}
  ,
\end{align}
where we have used $\beta = \beta_p$ at $t = T_p$.

For the investigation of the correlation, we mainly fit the data to
the relation
\begin{equation}
  \label{eq:luminosity_relation}
  \log L_{X, p} = a + b \log \frac{T_p}{1+z}
  ,
\end{equation}
where $T_p/(1+z)$ is the corresponding transition time in the burst
frame.
Like many other luminosity relations, this relation is by no means
an accurate one.
Due to the complexity of GRBs, it is hardly possible that
$\log L_{X, p}$ is fully determined by only $\log [T_p/(1+z)]$.
As usual, intrinsic scatter $\sigma_{\mathrm{int}}$ is introduced here,
i.e., lacking further knowledge, an extra variability that follows
the normal distribution $\mathcal{N}(0, \, \sigma_{\mathrm{int}}^2)$
is added to take into account the contributions to $\log L_{X, p}$
from hidden variables.
For the fit, we used the techniques presented
in~\citet{Agostini:2005fe}.
Let $x = \log [T_p/(1+z)]$ and $y = \log L_{X, p}$;
according to~\citet{Agostini:2005fe},
the joint likelihood function for the coefficients $a$ and $b$ and the
intrinsic scatter $\sigma_{\mathrm{int}}$ is
\begin{align}
  \label{eq:likelihood}
  \mathcal{L}(a, \, b, \, \sigma_{\mathrm{int}})
  &\propto
  \prod_i
  \frac{1}
  {
    \sqrt
    {
      \sigma_{\mathrm{int}}^2
      + \sigma_{y_i}^2
      + b^2 \sigma_{x_i}^2
    }
  }
  \nonumber \\
  &
  \times
  \exp
  \left[
    -
    \frac
    {
      \left(
        y_i - a - b x_i
      \right)^2
    }
    {
      2
      \left(
        \sigma_{\mathrm{int}}^2
        + \sigma_{y_i}^2
        + b^2 \sigma_{x_i}^2
      \right)
    }
  \right]
  ,
\end{align}
where $x_i$ and $y_i$ are corresponding observational data
for the $i$th GRB.

For GRB data, we used the samples compiled
in~\citet{Willingale:2006zh},
that is, the $107$ GRBs detected by both BAT and XRT on \emph{Swift}
up to 2006 August 1.
The $T_{90}$ duration for these $107$ GRBs were obtained from
\url{http://swift.gsfc.nasa.gov/}.
The uncertainties of the fitted parameters
in~\citet{Willingale:2006zh} are given in $90\%$ confidence level.
We symmetrize the errors and derive the corresponding $1 \sigma$
uncertainties by just dividing the $90\%$ confidence level errors by
$1.645$.
Unless stated explicitly, the errors in this paper are for the
$1 \sigma$ confidence level.
In our analysis, when not all of the GRBs have the needed parameters
available, we use the maximum subset of the $107$ GRBs satisfying the
requirement. For example, in Eq.~(\ref{eq:luminosity_Xp}), the
calculation of $L_{X, p}$ needs the observed redshift $z$, in addition
to $F_p$ and $\beta_p$, which are derived from the fit of GRB data.

\section{Results and discussion}

\begin{figure}[htbp]
  \centering
  \includegraphics[width = 0.45 \textwidth]{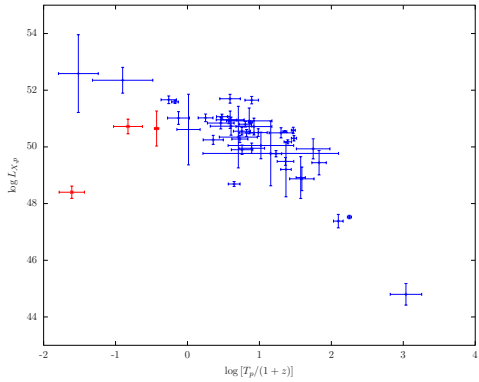}
  \caption
  {
    $L_{X, p}$ (in $\mathrm{erg} \, \mathrm{s}^{-1}$) versus
    $T_p/(1+z)$ (in seconds).
    $47$ GRBs are included with $37$ GRBs that have $T_p/(1+z)$
    greater than $2$ seconds and $3$ of which have $T_p/(1+z)$ greater
    than $100$ seconds.
    The red ones are conventional short GRBs
    ($T_{90} < 2 \, \mathrm{s}$).
  }
  \label{fig:Lp_vs_Tpz}
\end{figure}

\begin{figure}[htbp]
  \centering
  \includegraphics[width = 0.45 \textwidth]{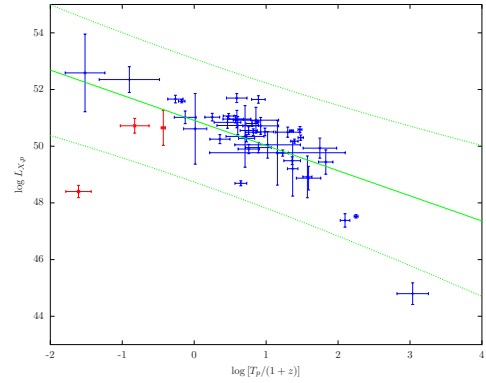}
  \\
  \includegraphics[width = 0.45 \textwidth]{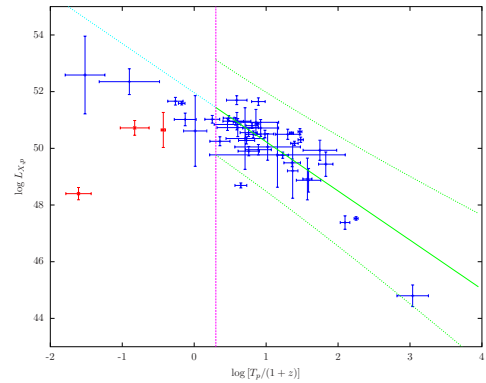}
  \caption
  {
    Best fit of the GRBs to the luminosity relation of
    Eq.~(\ref{eq:luminosity_relation}) and the corresponding
    $2 \sigma$ confidence region.
    Top: all GRBs are included in the fit.
    Bottom: only those with $T_p/(1+z) > 2 \, \mathrm{s}$ are
    included in the fit.
  }
  \label{fig:fit_result}
\end{figure}

\begin{figure}[htbp]
  \centering
  \includegraphics[width = 0.45 \textwidth]{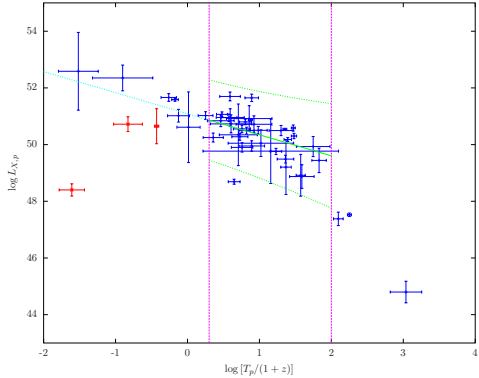}
  \caption
  {
    Best fit of the GRBs to the luminosity relation of
    Eq.~(\ref{eq:luminosity_relation}) and the corresponding
    $2 \sigma$ confidence region.
    Same as Figure~\ref{fig:fit_result} except that only GRBs with
    $2 \, \mathrm{s} < T_p/(1+z) < 100 \, \mathrm{s}$
    are included in the fit.
  }
  \label{fig:fit_result_subl}
\end{figure}

We plot the logarithm of $L_{X, p}$ versus the logarithm of
$T_p/(1+z)$ in Figure~\ref{fig:Lp_vs_Tpz}, which include $47$ GRBs.
We can see that most of the GRBs ($34$ GRBs) lie in the range of
$2 \, \mathrm{s} < T_p/(1+z) < 100 \, \mathrm{s}$.
There is obviously a correlation between $L_{X, p}$ and $T_p/(1+z)$.
However, when fitting the GRBs to the relations of
Eq.~(\ref{eq:luminosity_relation}), we have different options
depending on how we view the three GRBs with the largest $T_p/(1+z)$,
i.e., those with $T_p/(1+z) > 100 \, \mathrm{s}$.
For the first choice, we can simply include all the data points in the
fit (see the top panel of Figure~\ref{fig:fit_result}), which leads to
a result that the GRB with the largest $T_p/(1+z)$ lies outside the
$2 \sigma$ confidence region of the fit.
Alternatively, it is also possible that the GRBs with the largest
$T_p/(1+z)$, instead of just being outliers, may indeed reveal some
trend of the luminosity relation at large $T_p/(1+z)$.
In this case, we cannot simply ignore the GRBs with large $T_p/(1+z)$
just because their quantity is small and, if they are taken more
seriously, it seems that the samples are split into two groups at
some value of $T_p/(1+z)$ based on the slope of the luminosity
relation of Eq.~(\ref{eq:luminosity_relation}).
To show this, we perform a fit using only the GRBs with
$T_p/(1+z) > 2 \, \mathrm{s}$
(see the bottom panel of Figure~\ref{fig:fit_result}),
which gives a result quite different from the prior fit.
From the bottom panel of Figure~\ref{fig:fit_result}, we can see that
if the best fit line is extended to the range of
$T_p/(1+z) < 2 \, \mathrm{s}$,
all the GRBs with $T_p/(1+z) < 2 \, \mathrm{s}$ lie below the line.
As a comparison, we also fit the GRBs with
$2 \, \mathrm{s} < T_p/(1+z) < 100 \, \mathrm{s}$
(see Figure~\ref{fig:fit_result_subl}),
which show that the difference is indeed introduced by the three GRBs
with the largest $T_p/(1+z)$
when compared with the fit of GRBs with
$T_p/(1+z) > 2 \, \mathrm{s}$.
We tabulate the fit results in Table~\ref{tab:fit_result}.
From the table, we can see that the values of $b$ for the first two
cases are quite different.
In addition, it is also interesting to note that,
for the first case, the slope $b$ is close to the slope
($-0.74_{-0.19}^{+0.20}$) of a similar luminosity relation about
$T_a$ and luminosity at $T_a$ presented in~\citet{Dainotti:2008vw}.
For the second case, the slope $b$ is close to the index
($-1.5 \pm 0.16$) of the power-law declination of the average
luminosity of X-ray flares as a function of time presented
in~\citet{Lazzati:2008da}.
These coincidences may be worth noting in future studies with a bigger
sample of GRBs.

\begin{table}[htbp]
  \centering
  {\tiny
  \begin{tabular}{cccc}
    \hline \hline
    GRB set & $a$ & $b$ & $\sigma_{\mathrm{int}}$ \\
    \hline
    All available GRBs
    & $50.91 \pm 0.23$ & $-0.89 \pm 0.19$ & $1.06 \pm 0.13$
    \\
    $\frac{T_p}{1+z} > 2 \, \mathrm{s}$
    & $51.96 \pm 0.32$ & $-1.73 \pm 0.25$ & $0.78 \pm 0.11$
    \\
    $2 \, \mathrm{s} < \frac{T_p}{1+z} < 100 \, \mathrm{s}$
    & $51.09 \pm 0.32$ & $-0.74 \pm 0.30$ & $0.63 \pm 0.09$
    \\
    \hline
  \end{tabular}
  }
  \caption
  {
    Results of the fit to the luminosity relation of
    Eq.~(\ref{eq:luminosity_relation}).
  }
  \label{tab:fit_result}
\end{table}

Generally speaking, in statistical analysis, except for the different
measurement precisions, we should treat all the data points equally
and should not give more attention to some data points over the
others.
However, since our sample of GRBs here is very limited, and there are
only a few GRBs with very large $T_p/(1+z)$, some selection rules may
have already been imposed implicitly on the sample itself.
Considering this, it may be unfair to the GRBs with large $T_p/(1+z)$
to be treated as outliers just because their quantity is small,
especially when they may reveal some trend in the luminosity
relation.
This is why we consider both cases in the analysis,
i.e., whether the few GRBs with large $T_p/(1+z)$ should be treated
as just outliers or taken more seriously.
Correspondingly, the relation between $L_{X, p}$ and $T_p/(1+z)$ could
be a simple power law or a broken power law with a change in the slope
of Eq.~(\ref{eq:luminosity_relation}) at some characteristic value of
$T_p/(1+z)$.
For the present sample of GRBs used in our analysis, it is not
sufficient for us to conclude which case is real and, for the later
one, to determine the exact value of $T_p/(1+z)$ where the slope of
the luminosity relation changes.
Here, we leave it open to future studies with more GRBs and discuss
in the following the implications of the luminosity relation,
especially in the situation where there is a change in the
slope at some value of $T_p/(1+z)$.

\begin{figure}[htbp]
  \centering
  \includegraphics[width = 0.45 \textwidth]{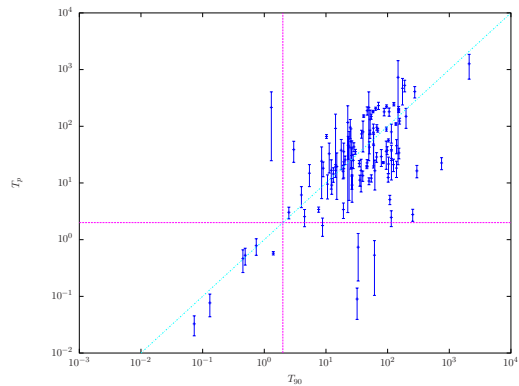}
  \caption
  {
    $T_p$ versus $T_{90}$ for the $107$ GRBs.
    Times are in seconds.
    On the cyan dash-dotted line the two quantities have equal values
    and
    the red dashed lines correspond to $T_p$ and $T_{90}$ equal to $2$
    seconds respectively.
  }
  \label{fig:Tp_vs_T90}
\end{figure}

\begin{figure}[htbp]
  \centering
  \includegraphics[width = 0.45 \textwidth]{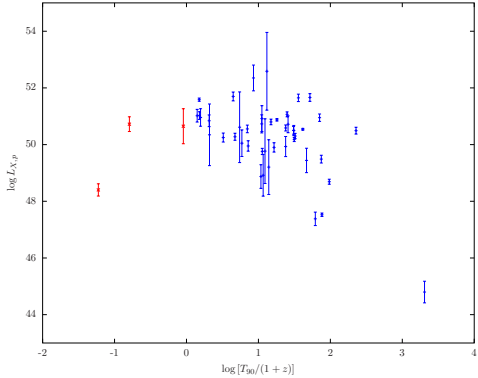}
  \caption
  {
    $L_{X, p}$ (in $\mathrm{erg} \, \mathrm{s}^{-1}$) versus
    $T_{90}/(1+z)$ (in seconds).
    The set of GRBs is the same as that in
    Figure~\ref{fig:Lp_vs_Tpz}.
    The red ones are conventional short GRBs
    ($T_{90} < 2 \, \mathrm{s}$).
  }
  \label{fig:Lp_vs_T90z}
\end{figure}

First of all, we emphasize that the luminosity relation is between the
luminosity and $T_p$,
though $T_p$ and $T_{90}$ can both act as an estimate of the GRB
duration and are comparable to each other for most GRBs, as can be
seen from Figure~\ref{fig:Tp_vs_T90}.
A similar relation seems not to exist between the luminosity and the
$T_{90}$. See Figure~\ref{fig:Lp_vs_T90z}; the corresponding data
points turn out to be very dispersive.
Despite the fact that most GRBs in Figure~\ref{fig:Tp_vs_T90} are
distributed around the line on which they are equal, there are some of
them that have been considerably different from $T_p$ and $T_{90}$.
In fact, the two quantities differ from each other significantly from
their derivation.
$T_{90}$ is calculated directly by using the BAT data in the
$15$--$150$
keV band, while for the calculation of $T_p$, the BAT data are first
extrapolated to the XRT band of $0.3$--$10$ keV in order to be
combined with the XRT data. For $T_{90}$, the emphasis is on the
percent of the fluence of a burst, while for $T_p$, the emphasis is on
the transition in a GRB light curve from the exponential decay in the
prompt phase to the initial power-law decay. In the time interval of
$T_{90}$, $90\%$ of the total fluence is observed, while the ratio
between the observed fluence in the time interval of $T_p$ and the
total fluence depends not only on the temporal decay index of the
initial power-law decay duration in the prompt component, but also on
the shape of the light curve in the afterglow component.
In addition, we must remember that large flares have been masked out
from the light curves before performing the fit that allows us to
derive $T_p$~\citep{Willingale:2006zh}.

As stated above, because the data are limited, there are two possible
models for the luminosity relation of
Eq.~(\ref{eq:luminosity_relation}), i.e.,
the luminosity relation, using one set of values for the parameters
(the intercept $a$ and the slope $b$), may be applicable to all the
GRBs except for some outliers.
or there may be a change in the slope $b$ of the luminosity relation
at some value of $T_p/(1+z)$.
If there is a change in slope this may suggest that GRBs could be
classified into two groups based on their values of $T_p/(1+z)$.
Since $T_p/(1+z)$ is an estimate of the GRB duration, this is in fact
an indication of how we should classify GRBs into long and short
ones and is actually the same as the proposal
by~\citet{O'Brien:2007apss} to use $T_p$ as the criterion for the
classification of long and short GRBs.
In principle, we should use the quantity in the burst frame
($T_p/(1+z)$) instead of that in the observer frame ($T_p$).
However, for a large portion of the observed GRBs, the redshifts are
not available.
Generally speaking, due to the diversity of GRBs, the classification
of a GRB is unlikely to be completely determined by only
its time duration, not to mention that the time duration of long and
short GRBs overlaps near the demarcation point.
Let us assume the demarcation point in the burst frame for the long
and short GRBs to be
\begin{equation}
  \label{eq:dempoint_burst}
  T_p/(1+z) = T_{l/s}
  .
\end{equation}
Then, as an approximate method, an effective redshift
$z_{\mathrm{eff}}$ can be defined, such that GRBs can be classified
into long and short ones in the observer frame at
\begin{equation}
  \label{eq:dempoint_observer}
  T_p = (1 + z_{\mathrm{eff}}) T_{l/s}
  .
\end{equation}
In addition, since a similar relation does not hold if we replace
$T_p$ with $T_{90}$ as mentioned previously, the change in the slope
$b$, which may be a reflection of different mechanisms, if confirmed,
would favor $T_p$ over $T_{90}$ as a criterion for the classification
of long and short GRBs.

\section{Summary}

We investigated the correlation between $T_p$ and the X-ray luminosity
of GRBs at $T_p$ and found a (broken) linear relation
between $\log L_{X, p}$ and $\log [T_p/(1+z)]$.
There may be a change in the slope of the relation at some value
of $T_p/(1+z)$ mainly because of the presence of the few GRBs with
large $T_p/(1+z)$.
The limited GRB sample used in our analysis is still not sufficient
for us to conclude whether the change in the slope is real or just an
illusion caused by outliers.
We considered both the cases in our analysis.
If the change is real, the different slopes may be a reflection
of different mechanisms for GRBs, which may suggest that using $T_p$
instead of $T_{90}$ (considering that a similar relation does not hold
if we replace $T_p$ with $T_{90}$, though $T_{90}$ and $T_p$ are both
estimates of the GRB duration) as a criterion for the classification
of long and short GRBs.

\acknowledgments

Shi Qi thanks Xue-Wen Liu, Lang Shao, Bo Yu, and Xue-Feng
Wu for helpful conversations.
We also thank the anonymous referee for many helpful suggestions and
comments.
We acknowledge the use of public data from the \emph{Swift} data
archive.
This research was supported by the National Natural Science Foundation
of China under grant No.~10973039 and the Jiangsu Planned Projects for
Postdoctoral Research Funds 0901059C (for Shi Qi).

\end{document}